\documentclass[11pt]{article}
\usepackage{graphicx}
\def\bx{\mathbf{x}}
\def\bX{\mathbf{X}}
\def\Nev{\mathcal{N}_{\rm ev}}

\begin{document}
\begin{center}
  \textbf{\Large Mathematics of complexity}\\
  \textbf{\Large in experimental high energy physics}\footnote{in:
    Second International Conference on Frontier Science: A Nonlinear
    World: The Real World, Pavia, Italy, 8-12 September 2003.
    Physica \textbf{A}338, 20-27 (2004).
  }\\[12pt]
  H.C.\ Eggers\\[12pt]
  \textit{Department of Physics, University of Stellenbosch,
    Stellenbosch, South Africa}\\[12pt]
\end{center}
\begin{abstract}
    Mathematical ideas and approaches common in complexity-related fields
    have been fruitfully applied in experimental high energy physics also.
    We briefly review some of the cross-pollination that is occurring.
\end{abstract}

Traditionally, the world of High Energy Physics (HEP) has not been
considered to be part of the loose conglomeration of topics
collectively termed \textit{complexity}.  While the topics and
communities are still largely disjoint, the degree and intensity of
overlap has grown as HEP has expanded from the study of single
particles (such as the top quark, the $W$-boson and the $B$-meson) to
precision measurements of complicated multiparticle systems.
Commonality and mutual inspiration between HEP and complexity is to be
sought not in the physics \textit{per se}, but in the underlying
mindset and approach. Specifically, the mathematics used in one may be
transmuted and adapted to the other to good effect. Indeed, it is the
mathematics and the mindset that unifies the diverse phenomena and
applications making up complexity.

This article seeks to highlight a few cases where such
cross-pollination has been occurring. We confine ourselves
specifically to \textit{experimental} high energy physics in the
knowledge that efforts to apply concepts and approaches of complexity
to theoretical HEP would widen the scope
considerably, \textit{cf.}~\cite{Cvi00a,Bir93a}.

\textbf{Events and their history:} A typical HEP experiment consists
of collecting a sample of many \textit{events}.  Each event occurs in
three phases: in the first, two incoming projectiles, which can be
leptons such as an electron and positron, hadrons such as protons or
mesons, or highly ionised nuclei, are made to collide at very high
energy. In Phase Two, the resulting ultrahigh concentration of energy
is converted, according to incompletely understood laws, into many
elementary particles which in turn collide with each other, transmute
and decay into lower-energy particles. In the third phase, the final
particles cease interacting and stream off into the detectors where
they are recorded.  Modern experiments typically accumulate millions
of events.

\textbf{Conditions and parameters specific to HEP:} 
For the purposes of comparing and contrasting the respective
mathematics, the following fundamental properties of HEP experimental
systems and data are of importance. 
\begin{enumerate}
  
\item[(A)] Recorded events are purely \textit{spatial} in character
  (with the ``phase space'' $\Omega$ defined e.g.\ in terms of
  momentum); there is no time ordering or time information. All
  mathematics relying on time ordering, such as time series analysis,
  is hence irrelevant from the start.  Also, all time information of
  the dynamics occurring in Phase Two of the collision can be inferred
  only indirectly from the available spatial information.
  
\item[(B)] Each recorded event is a \textit{point process}, i.e.\ it
  consists of $N$ structureless particles represented as points in the
  space. The multiplicity $N$ fluctuates from event to event. Not all
  particles are picked up by the detectors; considerable effort goes
  into correcting for this.
  
\item[(C)] There are many \textit{different types} of particles. In
  general, each particle is identified in terms of charge, mass, etc.,
  and its momentum is measured.
  
\item[(D)] Complex systems measurements and simulations commonly deal with
  $10^9$ data points. By contrast, each HEP event consists of
  \textit{very few particles}, ranging from a handful ($\sim 5$ for
  restricted measuring intervals or lower energies) to a maximum of
  $10^4$ in nucleus-nucleus collisions.
  
\item[(E)] While each event consists of comparatively few particles,
  \textit{huge samples} containing up to $10^7$ such events are now
  available. The net amount of data available in modern HEP
  experiments exceeds complexity data from measurement or simulation
  by orders of magnitude.
  
\item[(F)] The distribution of particles in the measuring space can be
  \textit{highly nonuniform in space}. Typical phenomena include the
  formation of strongly clustered ``jets'' in lepton-lepton and
  hadron-hadron collisions, kinematic effects, collective flow, and
  conservation laws.
  
\item[(G)] There can be differences from event to event even on a
  fundamental level due to \textit{uncontrollable or unknown
    parameters} such as the amount of overlap between colliding
  projectiles.

\end{enumerate}
In summary, HEP experimental data can be characterised mathematically as a
\textit{large sample of sparse multispecies point processes.}

In both HEP and experimental complexity, the aim of characterisation
quantities and techniques is to eliminate candidate theoretical
models purporting to explain the results.  What, then, can the HEP
experimentalist learn from complexity?  Two common dominants have
emerged: the concept of scale and the use of multivariate statistics.
While both topics have been long familiar in complexity, the
properties listed above result in HEP-specific limitations,
adaptations and opportunities, which we now examine.

Given its point process character, the natural complexity counterpart
of an HEP event is a strange attractor, where points $\bx(t)$ of the
dynamical map are plotted in the embedding space while discarding time
information. While initially the focus of characterisation had been on
purely geometric properties of attractors, the advent of the
correlation integral \cite{Gra83a} and multifractals \cite{Hal86a},
permitted a more complete description in terms of both support and
measure \cite{Ren70a}. Box multifractals, for example, are defined in
terms of moments $\rho_q = \sum_m p_m^q\; (q=1,2,\ldots)$ of relative
frequencies $p_m = n_m/N$ (with $n_m$ the number of particles in bin
$m$, and $N = \sum_m n_m$).  If $\rho_q$ turns out to exhibit
power-law behaviour as a function of scale (bin size) $\ell$,
$\rho_q(\ell) \sim \ell^{(1-q)D_q}$, then the set of constants $D_q$
are termed ``generalised dimensions'' or R\'enyi dimensions
\cite{Gra83a,Ren70a}.

The predilection in complexity towards scale invariance proved a
fruitful inspiration to HEP experiments: scaling of suitably
normalised moments was, indeed, found in many cases; see \cite{Wol96a}
for reviews. The strong anisotropy in behavior parallel and
perpendicular to the collision axis has also induced measurements of
Hurst exponents \cite{Wu93a}. Due to Property (D), however, a
prejudice towards scaling is not helpful.  Multifractals arise most
naturally in infinite-generation multiplicative cascades
\cite{Eve92a}, and so jets, the hierarchical cascade structures of
particle formation in lepton-lepton collisions, would be the best
candidates for scaling.  Typical jet multiplicities are rather low,
however, so that the scaling interval will necessarily be small.
Nucleus-nucleus collisions, on the other hand, produce large numbers
of particles, but these originate not from cascades but from a
semi-thermalised second phase which effectively destroys information
on particle histories and hence is not conducive to scaling.

\textbf{Factorials:} To deal with the low-multiplicity problem, Bia\l
as and Peschanski \cite{Bia86a}, noting that the dynamical variable
relevant to HEP cascades was usually continuous while the measured
particles were necessarily discrete, postulated that the transition
from continuous dynamics to a discrete number of particles would be a
poissonian fluctuation. Based on this assumption, they showed that the
\textit{factorial moments} of $n$ discrete particles, $\langle
n(n{-}1)\cdots(n{-}q{+}1)\rangle$, correspond the \textit{ordinary
  moments} $\langle x^q \rangle$ of their continuous ancestor $x$.
Lipa \cite{Lip89a} showed early on that an explicitly scaling
continuous model with a continuous-to-discrete last step does continue
to exhibit scaling even for low multiplicities.

\textbf{Multivariate statistics:} While scaling as such was an
important addition to the HEP vocabulary, its narrow applicability
necessitated a wider approach: scaling assumptions are permitted in
multivariate statistics, but not required. While the HEP community had
long made use of the latter to characterise e.g.\ two-particle
correlations \cite{Foa75a}, the advent of complexity-inspired thinking
has boosted its significance and sophistication considerably.

The \textbf{sample of eventwise measures} $\hat\rho_e,
e=1,\ldots,\Nev$ serves as the starting point. There is a direct
analogy\footnote{ There are obvious differences, the first being that
  for a finite number $\hat N$ of particles, a HEP event cannot
  accommodate the limit $N{\to}\,\infty$ needed to define the
  invariant measure.  Clearly, $\hat\rho_e$ will also differ radically
  from event to event, so there is no question of invariance on this
  level.  } between the invariant measure of the strange attractor,
made up of points at the $i$-th iterate of map $f(\bx)$, and the
measure of an event $e$ made up of $\hat N$ particles measured at
phase space points $\{\bX_i\}_{i=1}^{\hat N}$:
\begin{equation}
\hat\rho(\bx) = \lim_{N\to\infty}\;\;\frac{1}{N}\;
  \sum_{i=1}^N \delta(\bx - f^i(\bx)) 
\qquad \rm{vs.} \qquad
  \hat\rho_e(\bx) = \sum_{i=1}^{\hat N} \delta(\bx - \bX_i) \,.
\end{equation}
\textbf{Phase space integrals:} Observation coordinate $\bx$ and the
data $\bX_i$ live in an embedding space $\Omega$.  In the language of
eventwise measures, multifractals and indeed all correlation
statistics are easily seen to be integrals over subregions of products
of $\Omega$. An example of a one-dimensional case $\Omega = [0,L]$ is
illustrated in Fig.~1 for $q=2$. The original event measure
$\hat\rho_e$ of a typical event is visualised as $\hat N$ dots living
on the line $[0,L]$, shown in Fig.~1 below the squares.  A product of
the measure with itself,\footnote{The unequal sign, enforcing
  factorial counting, subtracts out points on the diagonals, i.e.\ 
  those points where a single particle is counted as ``a pair with
  itself''.}  
$\hat\rho_e(\bx_1,\bx_2) = \sum_{i_1 \neq i_2}^N \delta({\bx_1} -
X_{i_1})\;\delta({\bx_2} - X_{i_2}),$ is then a set of dots on the
product space $\Omega\otimes\Omega$ represented by the squares in the
Figure.  Each dot represents a \textit{particle pair}, and second
moments for a particular domain are found by counting the number of
dots in a given domain.

Testing dependence on scale in box moments then amounts to integrating
over bin $\ell = L/2^k, k=1,2,\ldots$, counting only pairs falling in
the string of $(L/2^k)^2$-sized squares along the diagonal (see
Fig.~1(a)--(c)). As illustrated in Fig.~1(d), the correlation integral
\cite{Gra83a} is nothing but a series of strips of varying width
$\sqrt{q}\,\ell$ parallel to the main diagonal.  Many other ``slices
of phase space'' have been defined, such as autocorrelations,
fixed-bin correlations and statistics based on void intervals in the
densities.
%%~~~~~~~~~~~~~~~~~~~~~~~~~~~~~~~~~~~~~~~~~~~~~~~~~~~~~~~~~~~~~~~~~~
\begin{figure}[htpb]
\parbox[c]{40mm}{
    \includegraphics[scale=0.16]{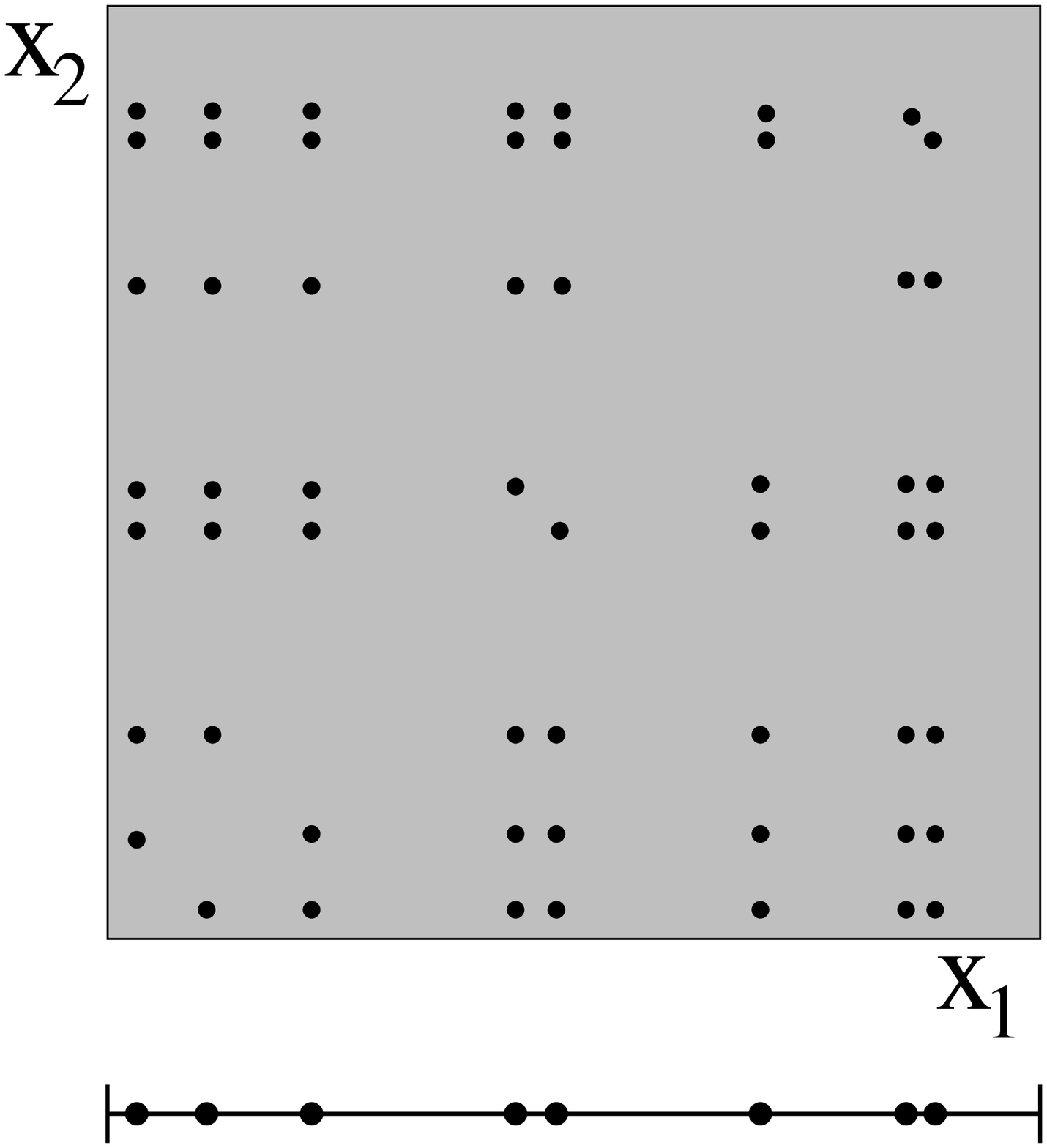}
}
\hspace*{-10mm}
\parbox[c]{40mm}{
    \includegraphics[scale=0.16]{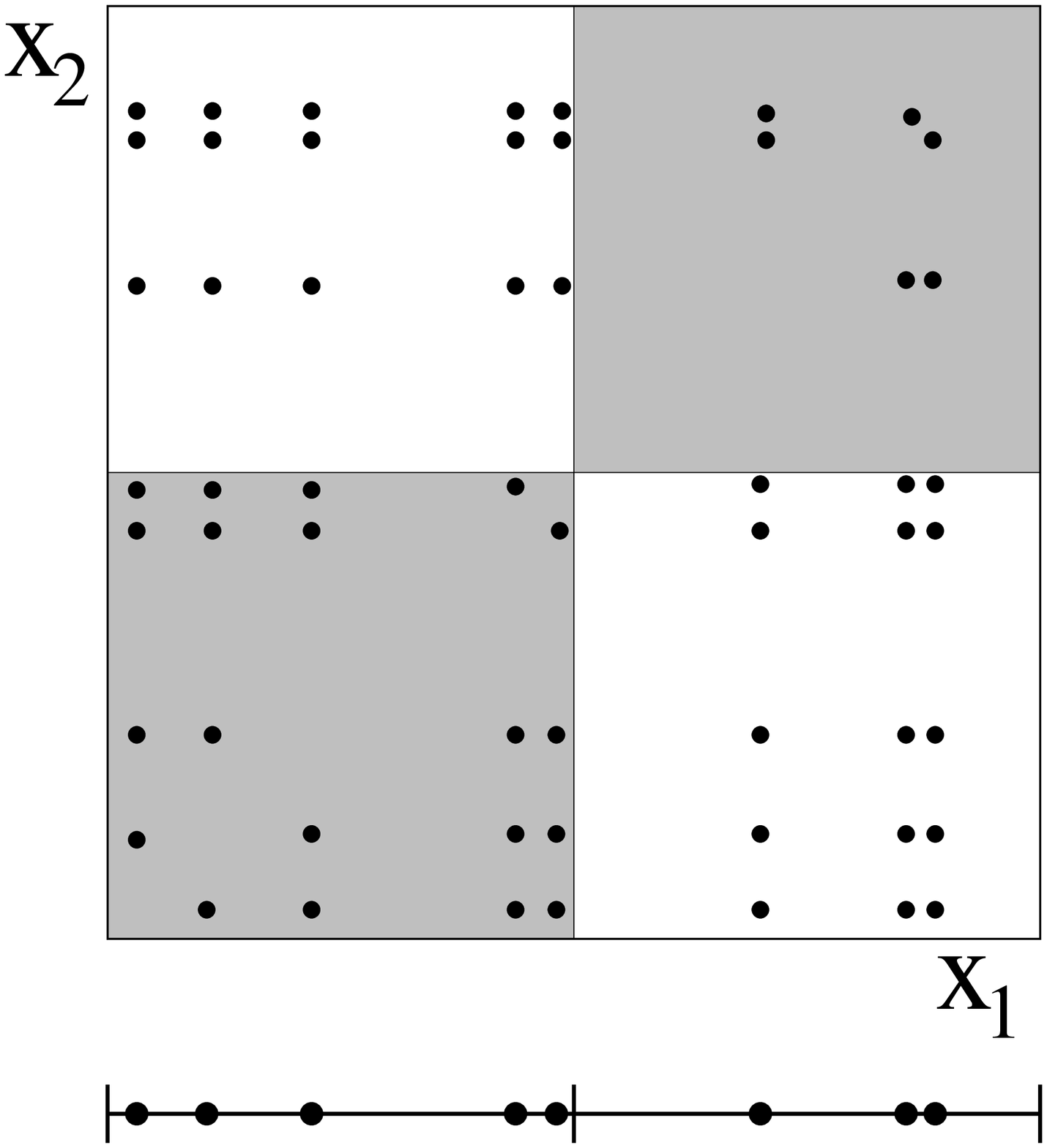}
}
\hspace*{-10mm}
\parbox[c]{40mm}{
    \includegraphics[scale=0.16]{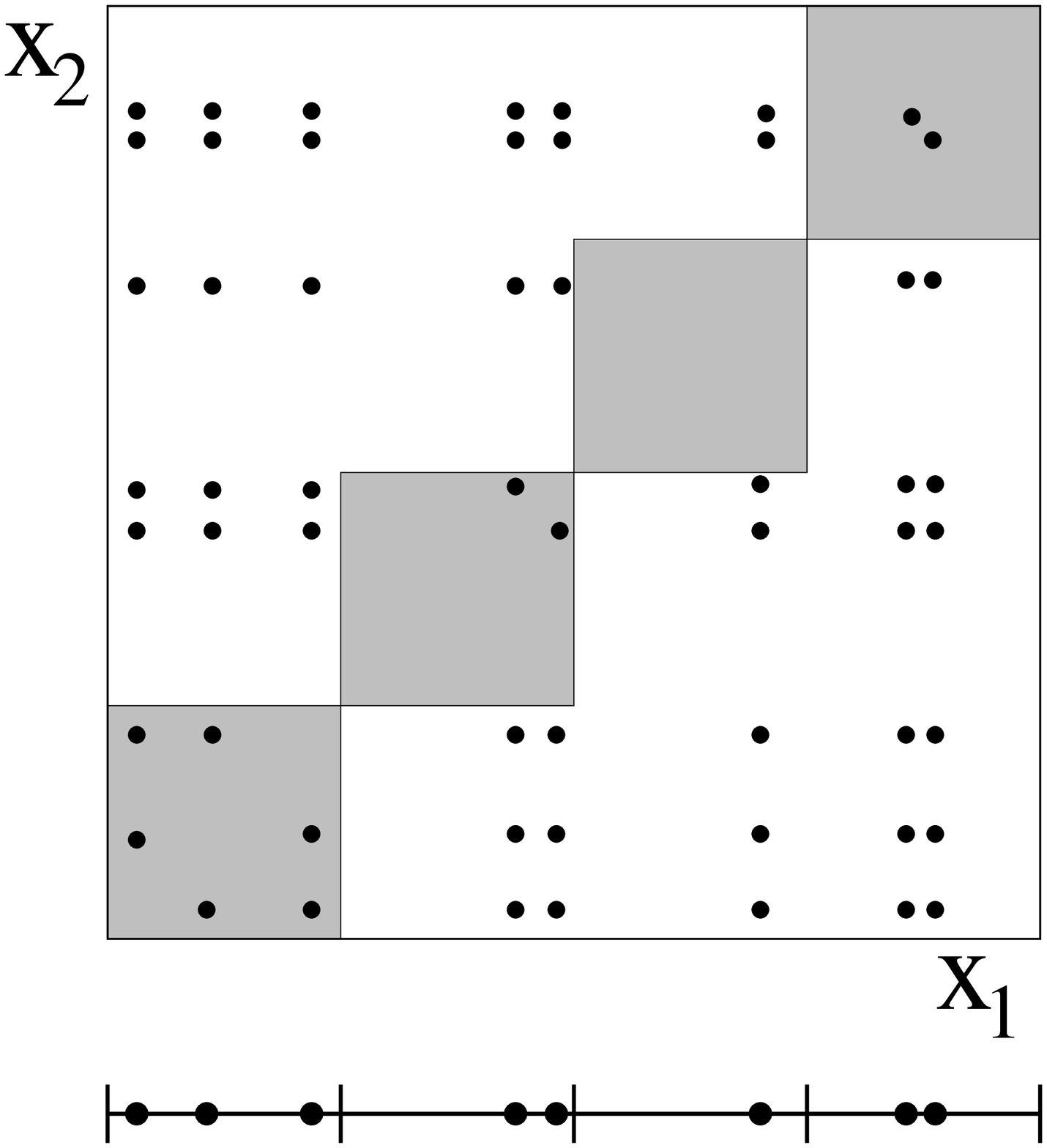}
}
\hspace*{-7mm}
\parbox[c]{40mm}{
    \includegraphics[scale=0.16]{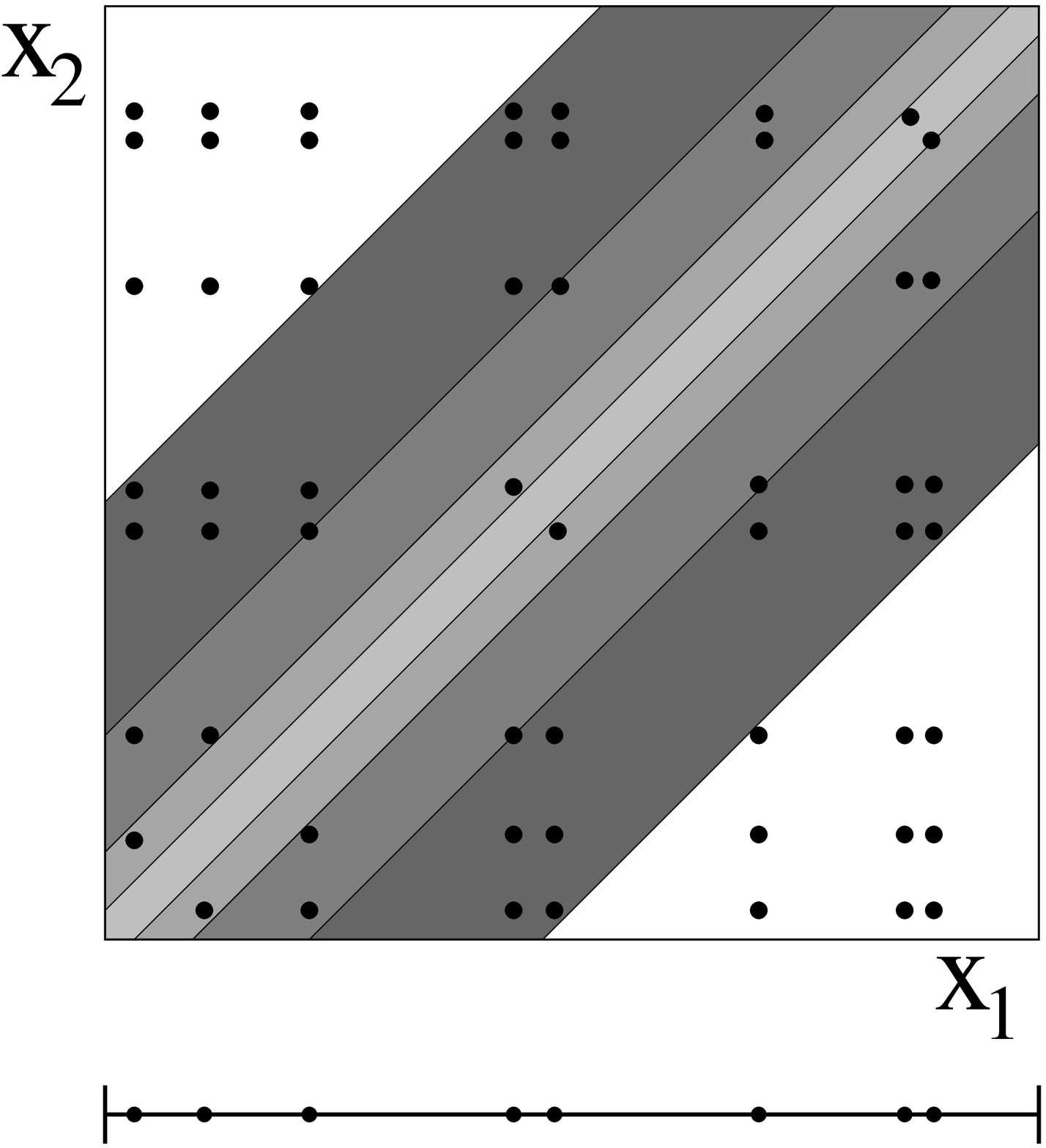}
} \\[-4mm]
{\small
\hspace*{14mm} (a) \hspace*{25mm} (b) \hspace*{25mm} (c) \hspace*{25mm} (d) 
}
\\[-5mm]
\caption{Representation of a single event in $\Omega$ (shown as dots on 
  the lines below the square) and in $\Omega\otimes\Omega$ space (in
  the squares).  (a)--(c): Domains of box moments as a function of
  scale, (d): corresponding correlation integral domains. Filling in
  the ``missing dots'' along the diagonals would change factorial to
  ordinary moments. }
%\end{center}
\end{figure}
%%~~~~~~~~~~~~~~~~~~~~~~~~~~~~~~~~~~~~~~~~~~~~~~~~~~~~~~~~~~~~~~~~~~

While higher orders $q{>}2$ are not visualised as easily, their
analysis proceeds analogously. Issues of ``topology'', i.e.\ the
combination of pairwise interparticle distances used to determine
$q$-tuple size, come into play \cite{Gra83a,Egg93a}.

Basing mathematical analysis on eventwise measures was instrumental in
deriving correlation integral prescriptions for a HEP context
\cite{Egg93a} and in providing a mathematical basis for \textit{event
  mixing}, whereby the uncorrelated background is simulated by
analysing artificial events made up of particles selected randomly
from different events in the sample.

%%%%%%%%%%%%%%%%%%%%%%%%%%%%%%%%%%%%%%%%%%%%%%%%%%%%%%%%%%%%%%%%%%%%%%%

\textbf{Cumulants:} The availability of large samples of
low-multiplicity events led to the direct measure of cumulants,
$\kappa_q$ \cite{Stu87a} which have proven to be central to later
experimental efforts due to their statistical properties and
sensitivity.\footnote{Their lowest-order forms, $\kappa_1 = \langle x
  \rangle =$ \textit{sample mean} and $\kappa_2 = \langle x^2 \rangle
  - \langle x \rangle^2 =$ \textit{sample variance} are, of course,
  universally known.} Their general properties include:
\begin{itemize}

  \item Cumulants are \textbf{zero} when there is no net correlation.
    
  \item Cumulants of the sum of independent random variables $\bx_i$ are
    \textit{additive},
    \begin{equation}
      \label{cumaddit}
      \kappa_q\bigl(\,{\textstyle\sum_i} \bx_i\, \bigr) = \sum_i \kappa_q(\bx_i)
    \qquad q = 1,2,3,\ldots
    \end{equation}
    
  \item The cumulant $\kappa_q(A-B)$ of \textit{distribution $A$
      relative to distribution $B$} is equal to the difference of the
    $A$- and $B$-distribution cumulants \cite{Lip96a},
    \begin{equation}
      \label{cumdif}
      \kappa_q(A-B) = \kappa_q(A) - \kappa_q(B) \,.
    \end{equation}
    $B$ may, for example, be a reference distribution and
    $\kappa_q(A-B)$ the measured deviations from this pre-defined reference.
    
  \item With the exception of $\kappa_1$, multivariate cumulants are
    \textit{tensors under affine transformations}; a simple example is
    $\kappa_q(c\bx) = c^q\,\kappa_q(\bx)$.
    
  \item Using \textit{event mixing}, cumulants can be calculated
    even for correlation integrals \cite{Egg93a}.

\end{itemize}
An example of the usefulness of cumulants is shown in Fig.~2: measured
second order cumulants for hadron-hadron collisions were fitted
using various parametrisations (only the power law does well), and the
same parametrisation then compared to measured third-order cumulants
\cite{Egg97a}. The discrepancy shows that the underlying model
assumptions fail on third order even while being successful for
$q{=}2$.
%%~~~~~~~~~~~~~~~~~~~~~~~~~~~~~~~~~~~~~~~~~~~~~~~~~~~~~~~~~~~~~~~~~~
\begin{figure}[htpb]
  \hspace*{8mm}
  \includegraphics[scale=0.35]{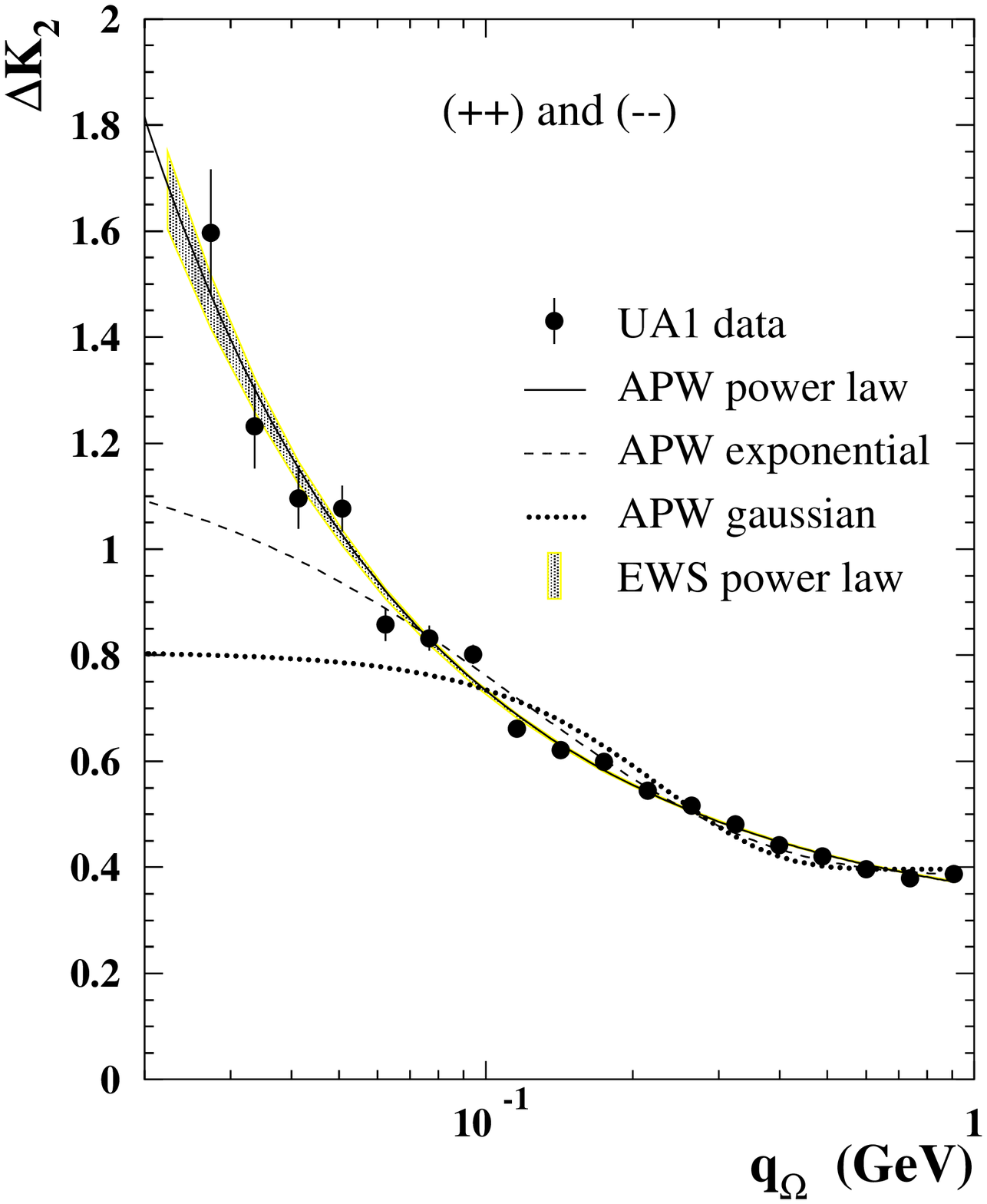}
  \hspace*{3mm}
  \includegraphics[scale=0.35]{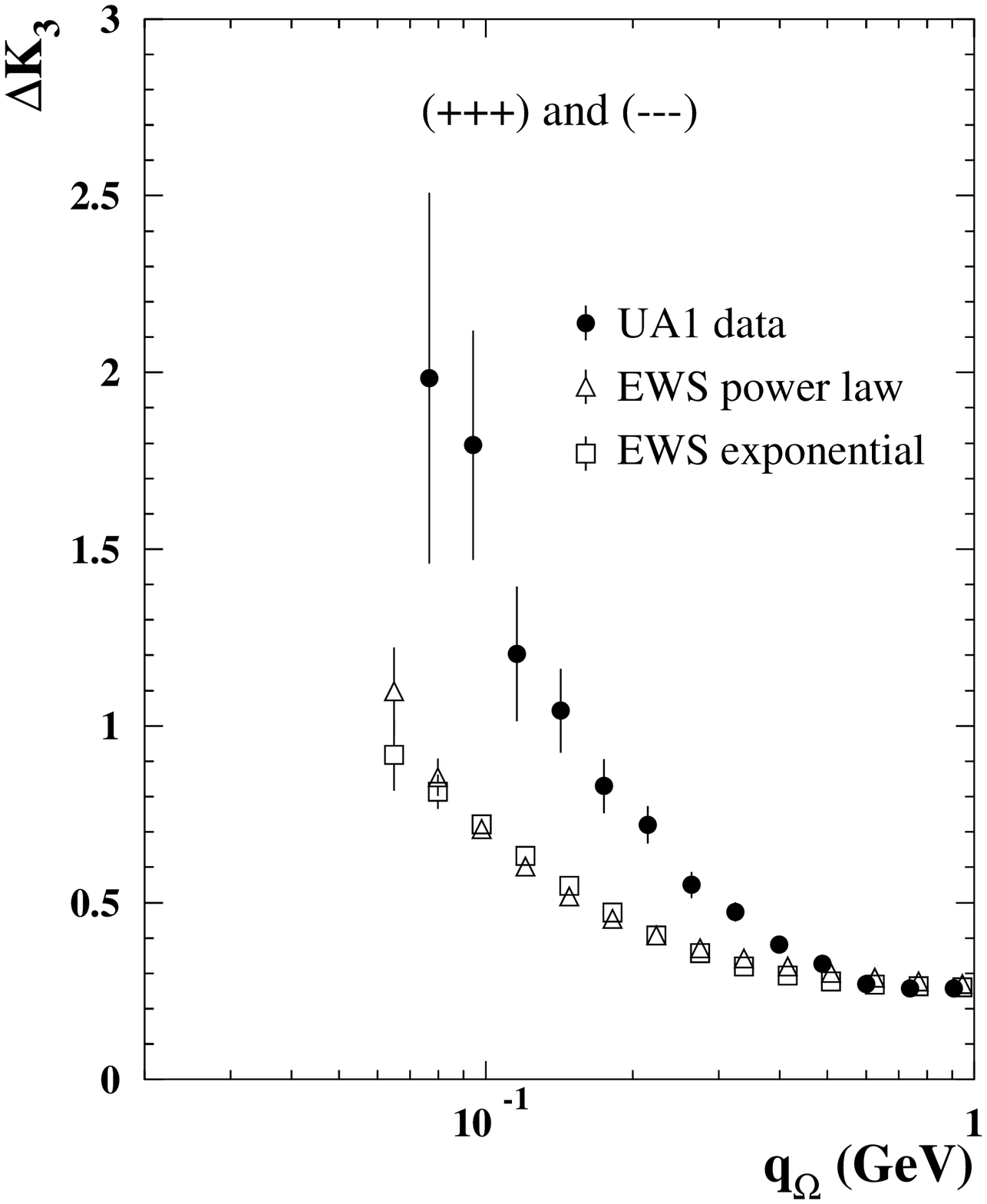}
  \caption{Use of cumulants in HEP to test theoretical models. Data was
    fitted to $q{=}2$ (left) and $q{=}3$ model predictions then
    compared to data (right).}
\end{figure}
%%~~~~~~~~~~~~~~~~~~~~~~~~~~~~~~~~~~~~~~~~~~~~~~~~~~~~~~~~~~~~~~~~~~

\textbf{Central Limit Theorem:} Capitalising on the many particles
generated in nucleus-nucleus collisions, recent experiments have
exploited the close link between the scale $\ell$ of a region and the
number $n$ of particles it can be expected to contain.  Assume,
simplistically, a constant number of particles $N$.  Given the
additivity property (\ref{cumaddit}), it is easy to show that the
cumulant of the average, $\langle x \rangle = N^{-1} \sum_{i=1}^N
x_i$, is equal to the cumulant over a smaller region containing but
one particle, suppressed by powers of $N$,
\begin{equation}
  \kappa_q\left(\langle x \rangle\right) 
  = N^{-q}\,  \sum_i\kappa_q(x_i) 
  = N^{1-q}\, \kappa_q(x) \,,
\end{equation}
\textit{if the $x_i$ are independent.}
Sensitive testing of independence is therefore provided by measuring
the deviation from zero of the statistic
\begin{equation}
  N^{q-1} \kappa_q\left(\langle x \rangle\right) - \kappa_q(x) \,.
\end{equation}
while varying the scale interval over which $\langle x\rangle$ is
calculated. This can be interpreted as an application of
Eq.~(\ref{cumdif}).

A more sophisticated version of this has recently been used
\cite{Sta03a} to partially address fluctuations in total eventwise
multiplicity also: With $p(m) = \sum_{i=1}^{n_m} p_i$ the sum of all
$n_m$ transverse momenta $p_i$ of particles found in bin $m$ of size
$\ell$ for \textit{one} event $e$, and $\overline p$ the average
over momenta of particles from \textit{all} events (``inclusive
average'') in the same bin, the \textit{total variance}, defined as
\begin{equation}
  \label{totvar}
  \Sigma_{p,n}^2(\ell) 
  = \overline{\sum_m \left( p(m) - n_m \overline{p} \right)^2} \,,
\end{equation}
turns out to be a difference between a $q{=}2$ cumulant over bins at
scale $\ell$ and a cumulant at the smallest available scale, thereby
effectively making use of Eq.~(\ref{cumdif}). Mul\-tiplicity
fluctuations are suppressed in Eq.~(\ref{totvar}) because
$\Sigma^2_{p,n}$ is based on bin-averaged versions of the standardised
variable $x^* = [x - \kappa_1(x)]/[\kappa_2(x)]^{1/2}$ which
automatically subtracts out mean multiplicity $\kappa_1$.

Fig.~3 shows schematically how $\Sigma^2_{p,n}$ can be expected to
behave as a function of scale $\ell \equiv \delta x$ \cite{Tra00a}.
Correlations manifest themselves as significant changes in
$\Sigma^2_{p,n}$ with a change in scale, while lack of correlation
shows up as invariance under change of measuring scale, which is
another formulation of the Central Limit Theorem (CLT).

%%~~~~~~~~~~~~~~~~~~~~~~~~~~~~~~~~~~~~~~~~~~~~~~~~~~~~~~~~~~~~~~~~~~
\begin{figure}[htpb]
\begin{center}
  \includegraphics[scale=0.38]{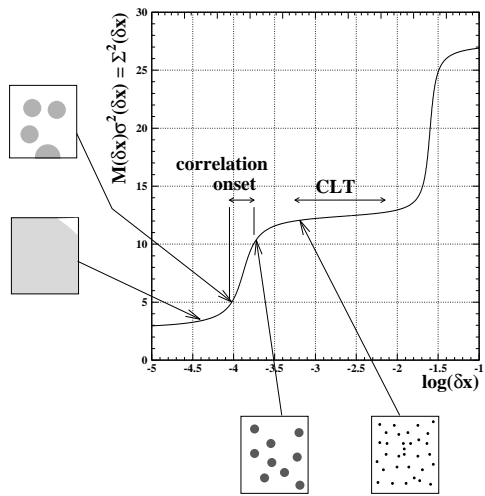}
  \vspace*{0mm}
  \caption{
    Changes with scale in correlation structure of ``points'' (shown under
    various magnifications in the little squares). Scale intervals of constant
    $\Sigma^2$ signify no correlation \cite{Tra00a}.}
  \end{center}
\end{figure}
%%~~~~~~~~~~~~~~~~~~~~~~~~~~~~~~~~~~~~~~~~~~~~~~~~~~~~~~~~~~~~~~~~~~

Other applications of cumulant differences include calculating
cumulants for a fixed-multiplicity sample using the multinomial as
reference distribution $B$ \cite{Lip96a}, and the calculation of
quantum mechanical interference effects between the jets resulting
from the weakly interacting bosons $W^+$ and $W^-$ \cite{L3-02a}.

\textbf{Event-by-event physics}: The availability of large samples
allows the study of very detailed special cases or events. Selecting
subsamples or plotting relative frequencies of events as a function of
some event property has become known as ``Event-by-event analysis''.
Some have attempted to characterise individual events, for example by
wavelet transforms \cite{Dre00a}, but so far, true event-by-event
analysis has been rare. Nevertheless, the degrees of freedom provided
by the sampling hierarchy \cite{Egg01a} of particles, events, sampling
distributions, special selections etc.\ are only just beginning to be
appreciated and exploited.
\\[2mm]

%%%%%%%%%%%%%%%%%%%%%%%%%%%%%%%%%%%%%%%%%%%%%%%%%%%%%%%%%%%%%%%%%%%%%%%%%%%%%
\textbf{Acknowledgments:} The author thanks the organisers for a 
well-organised and stimulating workshop. This work was supported in part
by the South African National Research Foundation.

%%%%%%%%%%%%%%%%%%%%%%%%%%%%%%%%%%%%%%%%%%%%%%%%%%%%%%%%%%%%%%%%%%%%%%%%%%%%%

%%%%%%%%%%%%%%%%%%%%%%%%%%%%%%%%%%%%%%%%%%%%%%%%%%%%%%%%%%%%%%%%%%%%%%%%%%%%%
%%%%%%%%%%%%%%%%%%%%%%%%%%%%%%%%%%%%%%%%%%%%%%%%%%%%%%%%%%%%%%%%%%%%%%%%%%%%%

\begin{thebibliography}{00}

\bibitem{Cvi00a}P.\ Cvitanovi\'c, Physica A\textbf{288}, 61 (2000);
                nlin.CD/0001034.

\bibitem{Bir93a}T.S.\ Biro, C.\ Gong, B.\ M\"uller and A.\ Trayanov,
                Int.\ J.\ Mod.\ Phys.\ C{5}, 113 (1994),
                nucl-th/9306002.

\bibitem{Gra83a}P.\ Grassberger and I.\ Procaccia,
                Phys.\ Rev.\ Lett.\ \textbf{50}, 346 (1983);\\
                P.\ Grassberger,
                Phys.\ Lett.\ A\textbf{97}, 227 (1983).

\bibitem{Hal86a}T.C.\ Halsey, M.H.\ Jensen, L.P.\ Kadanoff,
                I.\ Procaccia and B.I.\ Shraiman,
                Phys.\ Rev.\ A\textbf{33}, 1141 (1986).

\bibitem{Ren70a}A.\ R\'enyi, \textit{Probability Theory},
                North Holland (1970).

\bibitem{Wol96a}E.A.\ de Wolf, I.M.\ Dremin and W.\ Kittel,
                Phys.\ Rep.\ \textbf{270}, 1 (1996);
                %
                %\bibitem{Plo02a}
                M.\ Ploszajczak and R.\ Botet, 
               \textit{Universal Fluctuations: the
               phenomenology of hadronic matter},
                World Scientific (2002).

\bibitem{Wu93a}Y.\ Wu and L.\ Liu,
               Phys.\ Rev.\ Lett.\ \textbf{70}, 3197 (1993).

\bibitem{Eve92a}C.J.G.\ Evertsz and B.B.\ Mandelbrot, in:
               \textit{Chaos and Fractals}, H.O.\ Peitgen, 
                H.\ J\"urgens and D.\ Saupe, Springer (1992).

\bibitem{Bia86a}A.\ Bia\l as and R.\ Peschanski, 
                Nucl.\ Phys.\ B{\bf 273}, 703 (1986).

\bibitem{Lip89a}P.\ Lipa and B.\ Buschbeck, 
                Phys.\ Lett.\ B\textbf{223}, 465 (1989);
                P.\ Lipa, U.\ of Vienna PhD dissertation (1990)
                (unpublished).

\bibitem{Foa75a}L.\ Fo\` a, 
                Phys.\ Rep.\ \textbf{22}, 1 (1975).

\bibitem{Egg93a}H.C.\ Eggers, P.\ Lipa, P.\ Carruthers and B.\ Buschbeck, 
                Phys.\ Rev.\ D\textbf{48}, 2040 (1993).

\bibitem{Stu87a}A.\ Stuart and J.K.\ Ord, 
               \textit{Kendall's Advanced Theory of Statistics}, 
               Vol.1, Oxford University Press (1987).

\bibitem{Lip96a}P.\ Lipa, H.C.\ Eggers and B.\ Buschbeck,
                Phys.\ Rev.\ D\textbf{53}, R4711 (1996).

\bibitem{Egg97a}H.C.\ Eggers, P.\ Lipa, P.\ Carruthers, B.\ Buschbeck, 
                Phys.\ Rev.\ Lett.\ \textbf{79}, 197 (1997).

\bibitem{Sta03a}STAR Collaboration, J.\ Adams et al.,
                %Phys.\ Rev.\ Lett.\ 
                nucl-ex/0308033.

\bibitem{Tra00a}T.A.\ Trainor, hep-ph/0001148.

\bibitem{L3-02a}L3 Collaboration, P.\ Achard et al., 
                Phys.\ Lett.\ B\textbf{547}, 139 (2002).
                %CERN-EP/2002-062

\bibitem{Dre00a}I.M.\ Dremin et al., 
                Phys.\ Lett.\ B\textbf{499},97 (2001);
                hep-ph/0007060.

\bibitem{Egg01a}H.C.\ Eggers, 
                in: \textit{30th International Symposium on
                Multiparticle Dynamics}, Tihany, Hungary, 9--15 October
                2000, World Scientific (2001), pp.\ 291--302;\\
                hep-ex/0102005.

\end{thebibliography}
\end{document}